\documentclass[12pt]{article}
\usepackage{graphicx}

\def\be{\begin{equation}}
\def\ee{\end{equation}}

\def\Re{\mbox{\rm Re~}}
\def\Im{\mbox{\rm Im~}}

\def\ReNum{{\textsf{Re}}}

\begin{document}

\title{AC electrolyte conductivity in the $\omega\tau <1$ regime}

\author{I.Chikina$^1$, S.Nazin$^2$, and V.Shikin$^2$\\
%\shortauthor{I.Chikina, S.Nazin, and V.Shikin}
%
%\institute{
  $^1$DRECAM/SCM/LIONS CEA - Saclay, 91191\\ Gif-sur-Yvette, Cedex, France\\
  $^2$ISSP RAS, Chernogolovka, Moscow district, 142432 Russia}

%\pacs{61.20.Qg}{\textbf{First pacs description}}
%\pacs{66.10.-x}{Diffusion and ionic conduction in liquids}
%\pacs{66.10.Ed}{Ionic conduction}

\maketitle

\begin{abstract}Details of the dynamic behaviour of different
charged clusters in liquids are discussed. Their associated mass
is considered which possesses a number of interesting features in
a normal viscous liquid.
\end{abstract}

In the two previous papers [1, 2] the authors noted that the
systematically observed [3--6] temperature dependence of the
positive ions effective mass in superfluid helium has mainly
normal (non-superfluid) origin. Formally, the point is that under
non-stationary conditions the Stokes drag force $F(\omega)$ acting
on a sphere moving in a normal fluid actually has both real and
imaginary components. In the extreme case of low Reynolds numbers
one has [7, 8]
$$
     F\left(\omega\right)=6\pi\eta R\left(
    1+\frac{R} {\delta\left(  \omega\right) }\right)  v\left(
    \omega\right)+\nonumber
$$
$$
    3\pi R^{2}\sqrt{\frac{2\eta\rho}{\omega}}\left(
    1+\frac{2R}{9\delta\left(  \omega\right)  }\right)  i\omega
    v\left( \omega\right),
    \eqno(1)
$$
$$
    \delta\left( \omega\right) =\left( 2\eta/\rho\omega\right) ^{1/2}
$$
where $\rho$ is the liquid density, $\eta$ is its viscosity, $v$
is the sphere velocity,  $\delta\left( \omega\right)$ is the
so-called viscous penetration depth, and $R$ is the sphere radius
(or effective radius of the ion or a different particle or
cluster).

It is natural to identify the coefficient at $i\omega v(\omega)$
with the efficient associated mass of the cluster:
$$
     m^{ass}( \omega,R) =m_{id}(\rho,R )
     \left[1+\frac{9}{2}\frac{\delta(\omega)}{R}\right],
$$
$$
     m_{id}(\rho,R )=2\pi\rho R^3/3
     \eqno(2)
$$
The associated mass $m^{ass}$ proves to be frequency-dependent
(the dependence being rather strong at low frequencies) and this
circumstance should be actually taken into account when
considering the ion (cluster) dynamics employing the Navier-Stokes
equation.

Experiments [3--6] were carried out at finite frequencies, and the
data of Ref. [3] were obtained in the range of $\omega\tau <1$,
where $\tau$ is the the particle velocity relaxation time. Here,
Here, according to Eq. (2), the following asymptotics arising due
to the ion effective mass growth at low frequencies holds:
$$
    \frac{\Im v}{\Re v}\propto \omega^{1/2}, \eqno(3)
$$
As to the measurements reported in Ref. [3] they on the one hand
reveal a substantial polaronic effect (the ion mass proves to be
several times larger than $m^{id}$). On the other hand, at low
frequencies Eq. (3) does not hold and instead the limiting
behaviour
$$
    (\Im v/\Re v)_{\omega \to 0} \propto \omega,
     \eqno(3a)
$$
is observed which is typical of clusters with fixed mass. An
acceptable tradeoff between (2) and (3) is that the
$m^{eff}(\omega\to 0)\propto \omega^{-1/2}$ dependence should
reach a saturation in the vicinity of $\omega\tau \sim 1$, i.e. in
the frequency range where $\Re F(\omega)\simeq \Im F(\omega)$.
Bearing this in mind, it is easy to show [1] that the associated
mass $m_n^{ass}(\rho_n, R_n)$ for the superfluid helium transforms
to
$$
    m_n^{ass}(\rho_n, R_n)=2.1\pi R_n^3\rho_n,
     \eqno(4)
$$
which means that the mass becomes frequency independent and its
observed temperature dependence is due to the factor $\rho_n(T)$
in Eq. (4) and its maximum value substantially exceeds the ideal
associated mass $m_{id}$. However, the assumption $\Re
F(\omega)\simeq \Im F(\omega)$ is not quite consistent with the
conditions of the experiments reported in Ref. [3].

Actually the effective mass saturation in the limit
$m^{eff}(\omega\to 0)$ is related to the divergency in the linear
approximation of the integral
$$
    W=\int \rho(r)u^2(r)d^3r, \eqno(5)
$$
where $u(r)$ is the velocity field falling off anomalously slow
(as $r^{-1}$) at large distances from the body moving with a
constant velocity $v$ through the liquid. The details of this
non-linear scenario clearly indicated in Ref. [1] have not yet
been studied. It is the purpose of the present paper to consider
this scenario and discuss its applicability to data of Ref. [3].

1. Divergency of Eq. (5), and hence of the quantity $m^{ass}_{st}$
$$
    m^{ass}_{st}v^2/2=W, \eqno(5a)
$$
can be eliminated in the so-called Oseen approximation [1,7,8]
which reveals that the behaviour $u(r)\propto r^{-1}$ following
from the linearized Navier-Stokes equation and resulting into the
divergency (5) is actually replaced by the exponential decay of
the velocity field at distances $r>R/\ReNum$, $\ReNum=Rv/\eta \ll
1 $. Because of this exponential decay the quantity $m^{ass}_{st}$
becomes finite but acquires a non-linear dependence on velocity
$v$
$$
    m_{st}^{ass} \simeq \frac{m_{id}}{\ReNum}\ln{(1/\ReNum)},\qquad
    \ReNum=Rv/\eta\ll 1,
    \eqno(6)
$$
where $m_{id}$ is defined in Eq. (2) and \ReNum is the Reynolds
number. The additional factor in (6) accounts (with the
logarithmic accuracy) for the presence of a laminar trace
generated by a sphere moving through a viscous liquid.
\begin{figure}[tbp]%%%%%%%%%%%1
\begin{center}
\includegraphics*[width=8.5cm]{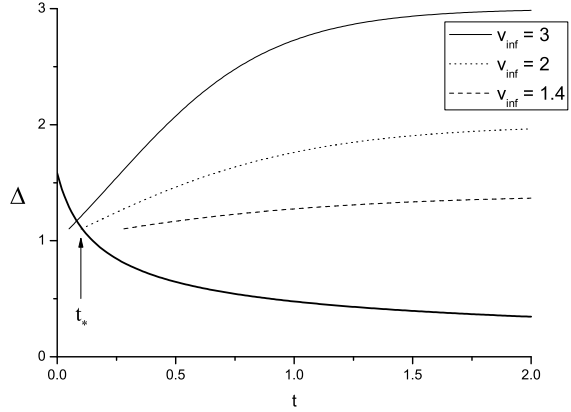}
\end{center}
\caption{Crossover from the $\Delta(t)$ curve defined by the
asymptotics (12) (thick solid line) to the long-time behaviour
described by Eq. (15). Thin solid, dotted, and dashed lines
correspond to $b/a=3, 2$ and 1.4, respectively.} \label{Fig_1}
\end{figure}
\begin{figure}[tbp]%%%%%%%%%%%3
\begin{center}
\includegraphics*[width=8.5cm]{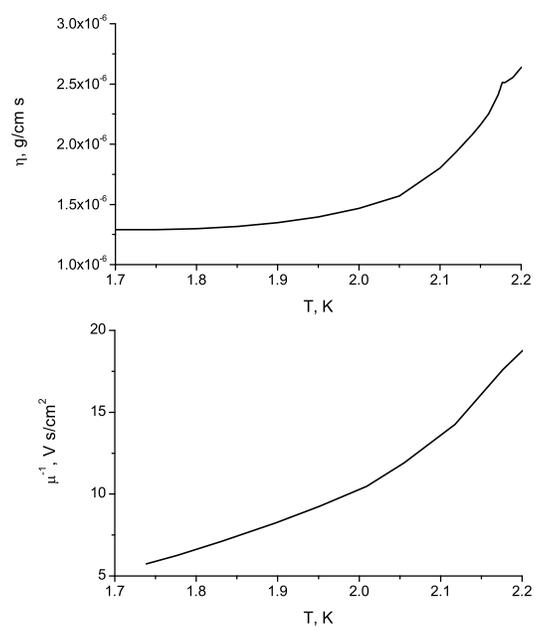}
\end{center}
\caption{Temperature dependence of the viscosity $\eta$ and
inverse mobility of negative ions $\mu^{-1}$ in helium.}
\label{Fig_2}
\end{figure}
\begin{figure}[tbp]%%%%%%%%%%%4
\begin{center}
\includegraphics*[width=8.5cm]{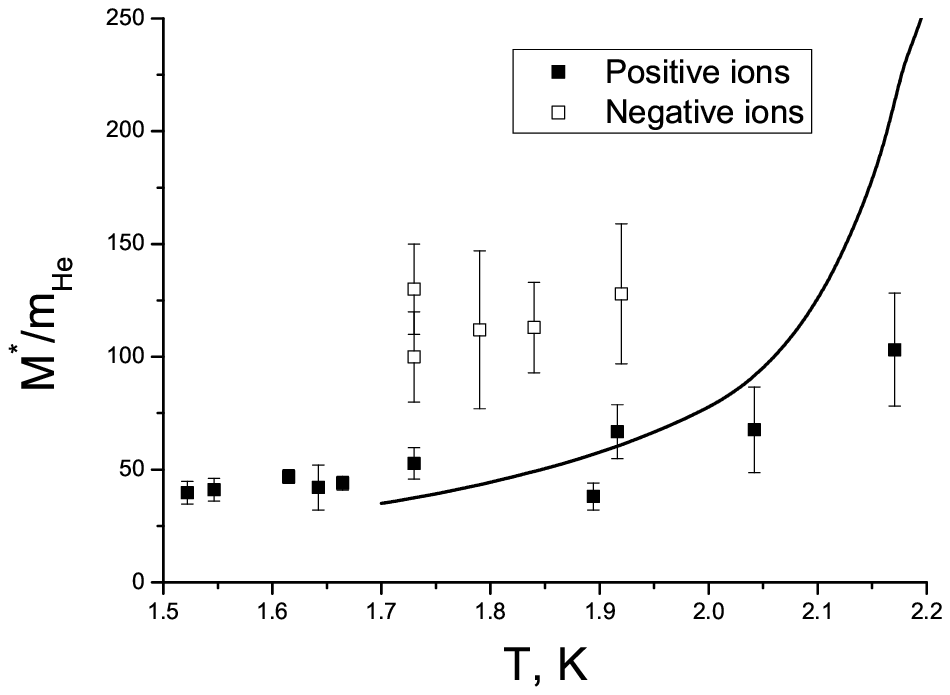}
\end{center}
\caption{Experimental (full squares) and theoretical (solid line
calculated according to Eq.(17)) effective mass of negative ions
in helium} \label{Fig_3}
\end{figure}

The result (6) is consistent with both the linear asymptotics
$m^{eff}(\omega\to 0)\propto \omega^{-1/2}$ and the requirement of
reaching a plateau in the limit. The same treatment reveals the
physical reasons of the divergency and the mechanism of its
elimination. However, the progress in understanding leads also to
some formal ``losses'' here since the cluster dynamics in the most
interesting domain $m_{st}^{ass}/m_{id}\gg 1$ becomes non-linear.
In particular, the Fourier representation providing a transparent
interpretation of the difference between (3) and (3a) can no
longer be used. Analysis of dynamic properties of the cluster
described by Eq. (6) requires an alternative approach. In the
present paper we discuss with this aim the problem of a step-like
external force $F$ acting on a particle in viscous liquid. An
appropriate quantity for the efficient particle mass is the
combination
$$
\Delta(t)=\frac{\dot{v}}{v_{\infty}-v(t)},
 \eqno(7)
$$
which is calculated below.

 For the Drude dynamics
$$
(m_0+m_{id})(\dot{v}+v/\tau)=eE(t), \eqno(8)
$$
the problem of a step-like external force $eE$ applied to the
particle leads to the velocity $v(t)$
$$
    v(t)=v_{\infty}[1-exp(-t/\tau)],\quad v_{\infty}=eE
    \tau/(m_0+m_{id}),
$$
$$
    \Delta_{Drude}=1/\tau,
    \eqno(9)
$$
where $m_0$ is the cluster bare mass, $m_{id}$ is
defined by Eq. (2), and  $e, E=const$ are the ion charge and
driving electric field.

On the other hand, the conventional Langevine equation
$$
(m_0+m_{id})\dot{v}+6\pi R \eta v(t)=eE(t) \eqno(10)
$$
yields
$$
    v(t)=v_{\infty}[1-\exp(-t/\tau_*)],\quad v_{\infty}=eE
    \tau_*/(m_0+m_{id}),\nonumber\\
$$
$$
    \tau_*^{-1}=\frac{6\pi R \eta}{(m_0+m_{id})}. \eqno(10a)
$$
$$
    \Delta_{Langev}=1/\tau_*
$$
In both cases (Eqs. (9) and (10)) the combination $\Delta$ does
not depend on time, and for the Drude case it also does not depend
on the ion mass.

It is natural to develop the Stokes dynamics which we are
interested in on the basis of the linear approximation where the
Fourier component $F(\omega)$ of the efficient drag force acting
on the sphere moving in a viscous liquid is given by Eq. (1). For
the velocity $v(t)$ this force yields [8,9]
$$
    v(t)=\frac{\gamma}{q}+\frac{\gamma p}{s_1^2-s_2^2}
    [\frac{\exp{(s_1^2t)}}{s_1}-\frac{\exp{(s_2^2t)}}{s_2}-
$$
$$
    \frac{1}{\sqrt{\pi}}\int_0^t\frac{\exp{[s_1^2(t-\tau)]}-\exp{[s_2^2(t-\tau)]}}{\sqrt{\tau}}
    d\tau ]\eqno(11)
$$
$$
    v(t=0)=0
$$
where $s_1$ and $s_2$ are roots of the equation
$$
    s^2+ps+q=0,
$$
$$
    q=\frac{\kappa}{m_0+m_{id}}, \quad
    p=3\frac{\sqrt{\kappa m_{id}}}
            {m_0+m_{id}},
$$
$$
    \gamma=\frac{eE}{m_0+m_{id}},\quad
    \kappa=6\pi R\eta.
$$
In agreement with the initial condition the left-hand side of Eq.
(11) turns into zero at $t \to 0 $ (which is readily verified if
one employs the formulae $s_1+s_2=-p$, $s_1s_2=q$).

In the opposite limiting case (where $|s^2_{1,2}|t\gg 1$) the
asymptotics
$$
    e^{s^2t}[\frac{1}{s}-\frac{1}{\sqrt{\pi}}
    \int_0^t\frac{\exp{(-s^2\tau)}}{\sqrt{\tau}}]\simeq
     \frac{1}{s^2\sqrt{\pi t}}[1+\sum_{n=1}^{\infty}
     \frac{n(2n-1)!!}{(2s^2t)^n}]
     \eqno(12)
$$
holds. The structure of Eq. (12) reveals that the velocity $v(t)$
approaches its asymptotic value $v(\infty)=\gamma/q$ following a
square-root law. It is also interesting that this asymptotics is
formed exponentially with the typical time
$$
\tau_{1,2}\sim s^{-2}_{1,2} \eqno(13)
$$

The appearance of exponentials (13) has a transparent qualitative
interpretation. At the initial stage of the adjustment of $v(t)$
to its steady-state value the process (11) resembles the Langevine
scenario (10) with the typical exponential relaxation to the
stationary behaviour and the relaxation time inversely
proportional to the constant cluster mass. Later, when the
associated mass starts to compete with the bare mass (either $m_0$
or $m_{id}$), a specific square-root approach to the stationary
regime develops which is absent in the traditional dynamics.

In addition to being of substantial interest in itself, the
results (11--13) proves to be important for correct formulation of
our main problem of the velocity relaxation for the ion with mass
(6). The point is that this definition does not apply at the
initial stage of the process where both $v(t)$ and the associated
mass are growing with time, i.e. $(d m/ d v)>0$ following the
scenario correctly described by Eq. (11). On the contrary, in the
situation described by Eq. (6) one has $(d m/ d v)< 0$.

Therefore, at some intermediate stage of the relaxation process
the quantity $(dm/dv)$ should have an extremum. At present we are
unable to determine its position on the time axis in a
self-consistent way, for example by solving the harmonic problem
in the Oseen approximation. Therefore, the suggested approximate
solution consists of two parts. The initial stage is described by
linearized dynamics (11--13). Its final part serves as the initial
condition for dynamics with mass (6). The matching time $t_*$ and
the corresponding ion velocity $v_*$ are taken to be $t_*\simeq
\tau_1$ and $v_*\simeq v(\tau_1)$ where $\tau_1$ is the longer one
of the two times $\tau_{1,2}$ (13).

The equation of motion to be solved is
$$
    m_*\dot{v}+\frac{m_{id}}{\ReNum\ln{(1/\ReNum)}}\dot{v} +6\pi R \eta
    v(t)=eE(t),
$$
$$
    m_*=m_0+ m_{id}\quad \ReNum=Rv/\eta \eqno(14)
$$
$$
v\ge v_* , \quad t\ge t_*,
$$
where $v_*$ and $t_*$ are the matching velocity and time. To find
$v_*$ and $t_*$, one should solve Eq. (14) for arbitrary values of
$v_*$ and $t_*$ and then to study the general conditions of the
intersection of curves $\Delta(t)$ resulting from Eqs. (11) and
(14). Then the intersection domain should be used to determine the
values of $v_*$ and $t_*$. In our approximate approach we adopt
$t_*\simeq \tau_1$, $v_*\simeq v(\tau_1)$.

In the most interesting limit $\ReNum\ll 1$ Eq. (14) can be
simplified and explicitly integrated to yield
$$
    \frac{1}{b}[\ln{\frac{v(t)}{b-av(t)}}-\ln{\frac{v_*}{b-av_*}}]=t-t_*,
$$
$$
    \quad a=\frac{6\pi R^2}{m_{id}}, \quad b=\frac{eER}{m_{id}\eta}.
\eqno(15)
$$
It is obvious that under the conditions $b-av(\infty)\to 0$ the
process (15) approaches the stationary regime with the velocity
$$
v(\infty)=v_{max}=eE/6\pi R\eta
 \eqno(16)
$$
The general behaviour of $\Delta(t)$ for $v(t)$ specified by
Eqs.(11) (in the range where asymptotics (12) is valid) and (15)
is presented in Fig. 1 where the choice of parameters $v_* $ and
$t_*$ is also illustrated.

 2. The above analysis allows two qualitative conclusions to be made.
First, in the domain $\omega\tau_* \ll 1$ the cluster associated
mass indeed has the structure (6) with the efficient velocity
$v=v_{max}$ reached by the ion in a single cycle. In these
estimates employing the results obtained for a semi-infinite time
interval the role of  $\omega^{-1}$ is played by appropriate
finite time interval containing the initial point. Second, in the
same frequency range $v_{max}$ can be estimated using either the
data on stationary ion mobility in the Stokes form (16), or direct
experimental measurements of that mobility covering also the
transitional (Knudsen-Stokes) domain.

Let us discuss the data of Ref. [3] within the framework provided
by Eqs. (14--16). First of all one should estimate the extent to
which the inequality $\omega\tau_* \ll 1$ is satisfied assuming
that $\omega \le 10^{-10}$ s and $\tau_*\ge \tau_1$ (13). The
available experimental data on the helium viscosity and normal
component density yield
$$
    \eta \simeq 2\cdot 10^{-5}\mbox{\rm\ g/(cm s)}, \qquad
    \tau_s =\rho_n R_s^2/2\eta\le 10^{-11} \mbox{\rm\ s},
$$
$$
    \omega\tau_s \sim  10^{-1}< 1.
$$

Further, the Reynolds number
$$
    \ReNum \sim \rho_n v_D R_s/\eta, \quad  \ReNum \ll 1
$$
proves to be dependent on the driving electric field strength
whose value was not given in Ref. [3]. It is only clear that лишь,
the field should be sufficiently weak so that the Reynolds number
is small (since otherwise it is impossible to explain the observed
mass enhancement). Under these conditions, bearing in mind that
$$
   l_{\eta}=\frac{R_s}{\ReNum},\quad M_{id}=2\pi R_s^3\rho_n/3
$$
and assuming that in all measurements the electric filed amplitude
$E_D$ was kept constant,
$$
    v_D=\mu E_D, \quad E_D=const
$$
one obtains
$$
    M_n^{ass} \simeq M_{id}\frac{l_{\eta}}{R_s}=
    \frac{2\pi R_s^2\eta}{3v_D}=
    \frac{2\pi R_s^2\eta}{3\mu E_D}
    =\rm{const}\cdot \frac{\eta}{\mu} \eqno(17)
$$
where $\mu$ is the stationary ion mobility [3]. Thus, in contrast
to Eq. (4) the arising interpretation of the efficient mass
temperature dependence is related to the simultaneous effects of
both $\eta(T)$ and $\mu(T)$. The corresponding plots together with
the data of Ref. [3] are presented in Figs. 2 and 3.

To sum up, one can say that the complicated temperature dependence
of the Stokes associated cation mass is a manifestation of a
rather general phenomenon inherent to motion of various mesoscopic
clusters through viscous liquid. In the extreme case of small
Reynolds numbers their efficient mass proves to possess a
substantial velocity dependence. The outlined effect should be
taken into account in the ion dynamics in various electrolytes as
well as in the calculations of equilibrium and dynamic properties
of colloid systems, etc.

This work was supported by the Program ``Physics of Condensed
Matter'' of the Presidium of Russian Academy of Sciences, and the
Russian Foundation for Basic Research. One of the authors (I.Ch.)
gratefully acknowledges support from ANR Grant No.
ANR-06-BLAN-0276.

\end{document}